\documentclass[aps,prl,showpacs,twocolumn,floats]{revtex4}
\usepackage{amssymb}
\usepackage{amsbsy}
\usepackage{amsmath}
\usepackage{graphicx}
\usepackage{amsmath}
\usepackage{times}
\usepackage{color}
\usepackage{subfigure}
\usepackage{setspace}
\usepackage{bm}% bold math
\newcommand\bea{\begin{eqnarray}}
\newcommand\eea{\end{eqnarray}}
\newcommand\beq{\begin{equation}}
\newcommand\eeq{\end{equation}}
\newcommand\bib{\bibitem}

\newcommand{\ket}[1]{| {#1} \rangle}
\newcommand{\expect}[1]{\langle {#1} \rangle}

\begin{document}

\title{Suppressing defect production during passage through a quantum critical point}

\author{Jay D. Sau$^1$ and K. Sengupta$^2$}
\affiliation{$^1$ Condensed Matter Theory Center, University of
Maryland, College Park, Maryland 20742, USA \\ $^2$ Theoretical
Physics Department, Indian Association for the Cultivation of
Science, Jadavpur, Kolkata-700032, India.}

\date{\today}

\begin{abstract}

We show that a closed quantum system driven through a quantum
critical point with two rates $\omega_1$ (which controls its
proximity to the quantum critical point) and $\omega_2$ (which
controls the dispersion of the low-energy quasiparticles at the
critical point) exhibits novel scaling laws for defect density $n$
and residual energy $Q$. We demonstrate suppression of both $n$ and
$Q$ with increasing $\omega_2$ leading to an alternate route to
achieving near-adiabaticity in a finite time for a quantum system
during its passage through a critical point. We provide an exact
solution for such dynamics with linear drive protocols applied to a
class of integrable models, supplement this solution with scaling
arguments applicable to generic many-body Hamiltonians, and discuss
specific models and experimental systems where our theory may be
tested.

\end{abstract}


\maketitle

The physics of closed quantum systems driven out of equilibrium has
received a lot of theoretical and experimental attention in recent
years \cite{pol1,dziar1,dutta1}. One of the central issues in this
field involves understanding excitation or defect production in such
a system upon its passage through a quantum critical point. It is
well known when such a passage occurs due to a slow linear quench of
a Hamiltonian parameter of the system characterized by a rate
$\omega$, the defect density $n$ and the residual energy $Q$ scale
with universal powers $n \sim \omega^{d \nu/(zv+1)},\, Q\sim
\omega^{(d+z)\nu/(z\nu +1)}$, where $d$ is the dimension of the
system and $z$ and $\nu$ are the dynamical critical and correlation
length exponents\cite{kz1,pol2}. Such scaling laws can also be
extended to cases where the system passes through a critical surface
\cite{ks1} and for non-linear ramps \cite{ks2,pol3,sondhi1}. These
scaling laws indicate an inevitable increase of $n$ with increasing
$\omega$.

Such an increase of $n$ and $Q$ is disadvantageous for the purpose
of quantum computation or dynamic preparation of a specific quantum
state which necessitate implementation of dynamical protocols taking
a quantum system from one ground state to another in a finite amount
of time. Consequently, several theoretical suggestions for
implementation of transitionless drive protocols have been put
forth. A class of such protocols involve modification of the system
Hamiltonian $H_0(t)$ by a suitably chosen $H_1(t)$ so that the
instantaneous ground state of $H_0(t)$ becomes the exact solution of
the many-body time dependent Schrodinger equation with
$H(t)=H_0(t)+H_1(t)$\cite{berry1,rice1}. Such a procedure has been
theoretically studied for several systems
\cite{polls1,polls2,berry1,dc1}. However, its experimental
implementation could be complicated; for example, for the transverse
field Ising model \cite{dc1}, $H_1(t)$ involves several multispin
non-local terms which might be difficult to implement in realistic
experimental systems. Another route to such nearly transitionless
dynamics involves use of optimal protocols as demonstrated for 1D
Luttinger models in Ref.\ \cite{rah1}. However, implementation of
these protocols for arbitrary many-body systems remains a challenge.

In this letter, we provide an alternative route to suppression of
defect density on passage through a quantum critical point. Our
method involves driving two parameters of the generic Hamiltonian
which reaches the critical point at $\lambda=\lambda_c \ne 0$ and
has a quasiparticle dispersion $E_{\bf k} = c |{\bf k}|^z \equiv c
k^z$ at the critical point. The first parameter driven according to
the protocol $\lambda(t)= (\omega_1 t-\lambda_c)^{\alpha}$, where
$\omega_1$ is the rate and $\alpha$ is a positive exponent, controls
the proximity of the system to the quantum critical point, while the
second $c(t)=|\omega_2 t|^{\beta}$ controls the dispersion of the
quasiparticles at the critical point. We show that under such a
drive, for $\omega_1/\lambda_c^{\alpha z \nu+1} \ll 1$ and
$\omega_2/\omega_1 \ge \omega_1^{\alpha \nu/[\beta(\alpha z \nu
+1)]}/\lambda_c$, the defect density $n$ scales as
\begin{eqnarray}
n \sim \omega_1^{\left(\frac{\alpha \nu}{\alpha z \nu +1} +
\frac{\beta}{z}\right) d} \omega_2^{-\beta d/z}  \label{scaling1}
\end{eqnarray}
leading to its suppression with increasing $\omega_2$. The scaling
of $Q$ is obtained by replacing $d \to (d+z)$ in Eq.\ \ref{scaling1}
and shows an analogous suppression. We note that our results
reproduce the standard single parameter scaling results of $n$ and
$Q$ \cite{kz1,pol2,ks2,pol3} as a special case for $\beta=0$ where
$c(t)$ becomes a time independent constant. We provide an exact
solution for a class of $d$-dimensional integrable models with
linear ramp protocols ($\alpha=\beta=1$) showing such behavior and
supplement it with scaling arguments leading to Eq.\ \ref{scaling1}
for arbitrary $\alpha$ and $\beta$. We also demonstrate a crossover
between regimes where $n$ and $Q$ increases/decreases with
increasing $\omega_{1}$ and $\omega_2$ with $\omega_2 =\omega_1^r$
($r>0$) and identity the exponent $r^{\ast}=1+ \alpha z
\nu/[\beta(\alpha z \nu+1)]$ at which the crossover occurs. Finally,
we discuss specific models and realistic experiments which could
provide a test for our theory. We note that our results constitute a
generalization of the well-known scaling laws for $n$ and $Q$
\cite{kz1,pol2,ks2,pol3}; in addition, they also provide a novel
route to achieving near-adiabatic drive protocols for taking a
quantum system through a critical point in finite time. We therefore
expect our work to be of interest to theorists and experimentalists
studying protocols of bit manipulations for quantum computation,
dynamic preparation of quantum states, and non-equilibrium dynamics
of strongly correlated quantum systems.

We begin by studying a class of $d$-dimensional integrable models
with a Hamiltonian $H(t) = \sum_{\bf k} \psi_{\bf k}^{\dagger}
H_{\bf k}(t) \psi_{\bf k}$, where $\psi_{\bf k}= (c_{1 \bf k},c_{2
\bf k})$ are Fermionic operators and $H_{\bf k}(t)$ is given by
\begin{eqnarray}
H_{\bf k}(t) = \tau_3 (\lambda_1(t) -b_{\bf k}) + \tau_1
\lambda_2(t) g_{\bf k}. \label{ham1}
\end{eqnarray}
Here $\tau_3$ and $\tau_1$ denote usual Pauli matrices while $b_{\bf
k}$ and $g_{\bf k}$ are general functions of momenta. We shall first
consider linear ramp protocol so that $\lambda(t)= \lambda_0
\omega_1 t$ and $\lambda_2(t)= \lambda_0 \omega_2 t$. In the rest of
this work, we shall set $\hbar=1$; all energy/frequency (time) units
shall be understood to be in units of $\lambda_0(\lambda_0^{-1})$.
The instantaneous eigenvalues of the Hamiltonian is given by $E_{\bf
k}(t) = \pm \sqrt{(\lambda_1(t) -b_{\bf k})^2+(\lambda_2(t) g_{\bf
k})^2}$. The critical point is reached at $t=t_{0{\bf k}_0}= b_{{\bf
k}_0}/\omega_1$ and ${\bf k}= {\bf k}_0$ where $g_{{\bf k}_0}=0$ and
$c(t)=|\lambda_2(t)|$ which reduces to $c(t_{\bf k_0})=|\omega_2
b_{\bf k_0}/\omega_1|$ at the critical point. In what follows, we
are going to assume that $b_{{\bf k}_0} \ne 0$ so that the critical
point is reached at a finite $t_{0{\bf k}_0} \ne 0$, and $g_{\bf k}
\sim |{\bf k}-{\bf k}_0|=k$ near the critical point.

\begin{figure}[t!]
\includegraphics[width=0.5\textwidth]{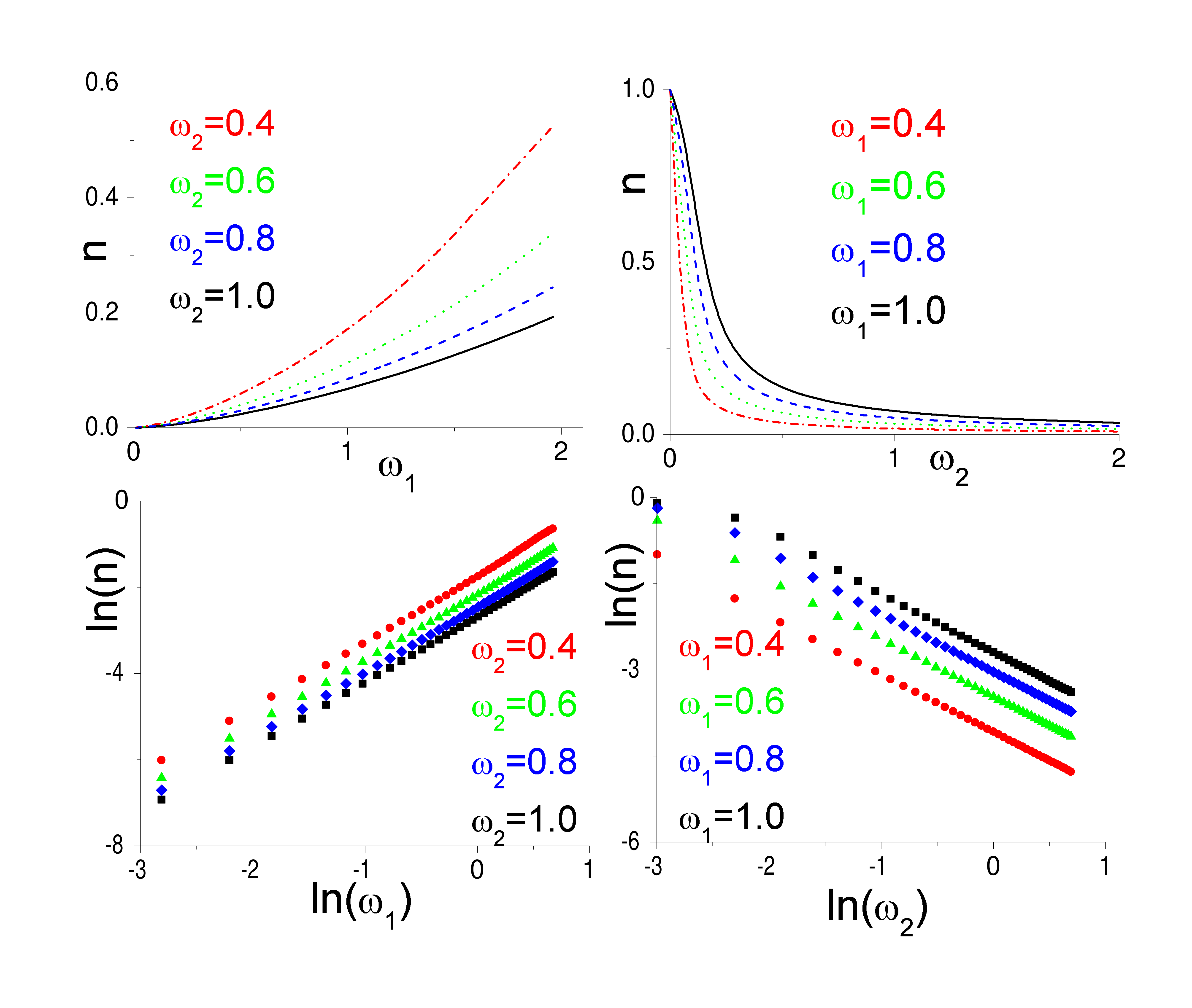}
\caption{(Color online) Top Panel: Plot of $n$ vs $\omega_1$ (left)
and $\omega_2$ (right) showing scaling of $n$. Bottom Panel: Plot of
$\ln(n)$ as a function of $\ln(\omega_1)$ (left) and $\ln(\omega_2)$
(right). All plots are computed using Eq.\ \ref{exexp} with $d=1$,
$b_{\bf k}=5-\cos(k)$, and $g(k)=\sin(k)$ so that $H$ represents 1D
XY model in a transverse field. The scaling regime, where Eq.\
\ref{scaling2} holds, occur for $\omega_2 \ge \omega_1^{3/2}/b_{\bf
k_0} =0.25 \omega_1^{3/2}.$ }\label{fig1}
\end{figure}

To obtain a solution for the dynamics we note that the Hamiltonian
density $H_{\bf k}(t)$ can be written in terms of a set of new Pauli
matrices ${\tilde \tau}_3$ and ${\tilde \tau}_1$ as
\begin{eqnarray}
H_{\bf k}(t) &=& \lambda_{1{\bf k}} (t-t_{1{\bf k}}) {\tilde \tau_3}
+ \lambda_{2{\bf k}} {\tilde \tau}_1,  \label{ham2}
\end{eqnarray}
where $t_{1{\bf k}} = b_{\bf k} \omega_1/\lambda_{1{\bf k}}$. In the
above expression, the quantities $\lambda_{1{\bf k}}$ and
$\lambda_{2{\bf k}}$ are given by
\begin{eqnarray}
\lambda_{1{\bf k}} &=& \sqrt{\omega_1^2 + \omega_2^2 g_{\bf k}^2}
\nonumber\\
\lambda_{2{\bf k}} &=& \sqrt{(\omega_2 t_{1{\bf k}} - b_{\bf
k})^2+\omega_2^2 g_{\bf k}^2 t_{1{\bf k}}^2}, \label{exact1}
\end{eqnarray}
so that $t_{1{\bf k}_0} = t_{0{\bf k_0}}$ and the matrices ${\tilde
\tau}_{1,3}$ can be expressed in terms of $\tau_{1,3}$ as
\begin{eqnarray}
\lambda_{1{\bf k}} {\tilde \tau}_3 &=& \omega_1 \tau_3 + \omega_2
g_{\bf k} \tau_1,  \nonumber\\
\lambda_{2{\bf k}} {\tilde \tau}_1 &=& (\omega_1 t_{1{\bf k}}-b_{\bf
k}) \tau_3 + \omega_2 t_{1{\bf k}} g_{\bf k}  \tau_1 \label{exact2}
\end{eqnarray}
We note that the above transformation transfer the entire time
dependence of $H_{\bf k}(t)$ to diagonal terms. From the structure
of $H_{\bf k}(t)$ (Eq.\ \ref{ham2}), it is easy to see that the
solution of the Schrodinger equation $i \partial_t \psi_{\bf k} =
H_{\bf k}(t) \psi_{\bf k}$ amounts to solving a Landau-Zener problem
for each ${\bf k}$ \cite{lz1}. For a linear ramp protocol where the
dynamics starts [ends] at $t \to -\infty [\infty]$, the probability
of defect production for any ${\bf k}$ can be simply read off as
\cite{lz1,vitanov1}
\begin{eqnarray}
p_{\bf k} &=& e^{- \pi \lambda_{2{\bf k}}^2 g_{\bf
k}^2/\lambda_{1{\bf k}}} = e^{- \pi \omega_{2}^2 b_{\bf k}^2 g_{\bf
k}^2/(\omega_1^2+\omega_2^2 g_{\bf k}^2)^{3/2}} \label{lzprob}
\end{eqnarray}
which leads to the defect density and residual energies to be
\begin{eqnarray}
n[Q] &=& \int d^dk /(2 \pi)^d 1[E_{\bf k}] e^{- \pi \omega_{2}^2
b_{\bf k}^2 g_{\bf k}^2/(\omega_1^2+\omega_2^2 g_{\bf k}^2)^{3/2}}.
\label{exexp}
\end{eqnarray}
For $\omega_{1,2}/b_{\bf k_0}^2 \ll 1 $ and $\omega_2/\omega_1 \ge
\sqrt{\omega_1}/b_{\bf k_0}$, $p_{\bf k}$ is appreciable around
${\bf k}= {\bf k}_0$, where $b_{\bf k}=b_{{\bf k}_0}$ so that
$p_{\bf k}= e^{-c\,\omega_2^2 k^{2}/\omega_1^3}$, with $c= \pi
b_{{\bf k}_0}^2$. Substituting the expression for $p_{\bf k}$ in
Eq.\ \ref{exexp} and rescaling $k' = k \omega_2/\omega_1^{3/2}$, one
obtains
\begin{eqnarray}
n \sim \omega_1^{3d/2} \omega_2^{-d}, \quad Q \sim
\omega_1^{3(d+1)/2} \omega_2^{-(d+1)}, \label{scaling2}
\end{eqnarray}
where we have used the fact $E_{\bf k} \sim k$ around ${\bf k}={\bf
k}_0$. Note that the scaling relations allows for large values
$\omega_2/\omega_1$; thus one can efficiently suppress defects by
tuning $\omega_2$ for a suitably chosen $\omega_1$. A plot of $n$
computed from Eq.\ \ref{exexp} with $d=1$, $b_{\bf k}= 5-\cos(k)$,
and $g_{\bf k}= \sin(k)$ (chosen so that the model conforms to 1D XY
model in a transverse field) is shown in top panels of Fig.\
\ref{fig1} as a function of the rates $\omega_1$ and $\omega_2$. The
plot clearly demonstrates that $n$ is a decreasing function of
$\omega_2$. In the scaling regime, the lines for different
$\omega_2(\omega_1)$ in the bottom left(right) panels of Fig.\
\ref{fig1} are parallel; their slope is numerically found to be
$1.507(-0.994)$ which agrees well with the theoretically predicted
values $3/2 (-1)$.

\begin{figure}[tbp]
\begin{center}
\includegraphics[width=0.5\textwidth]{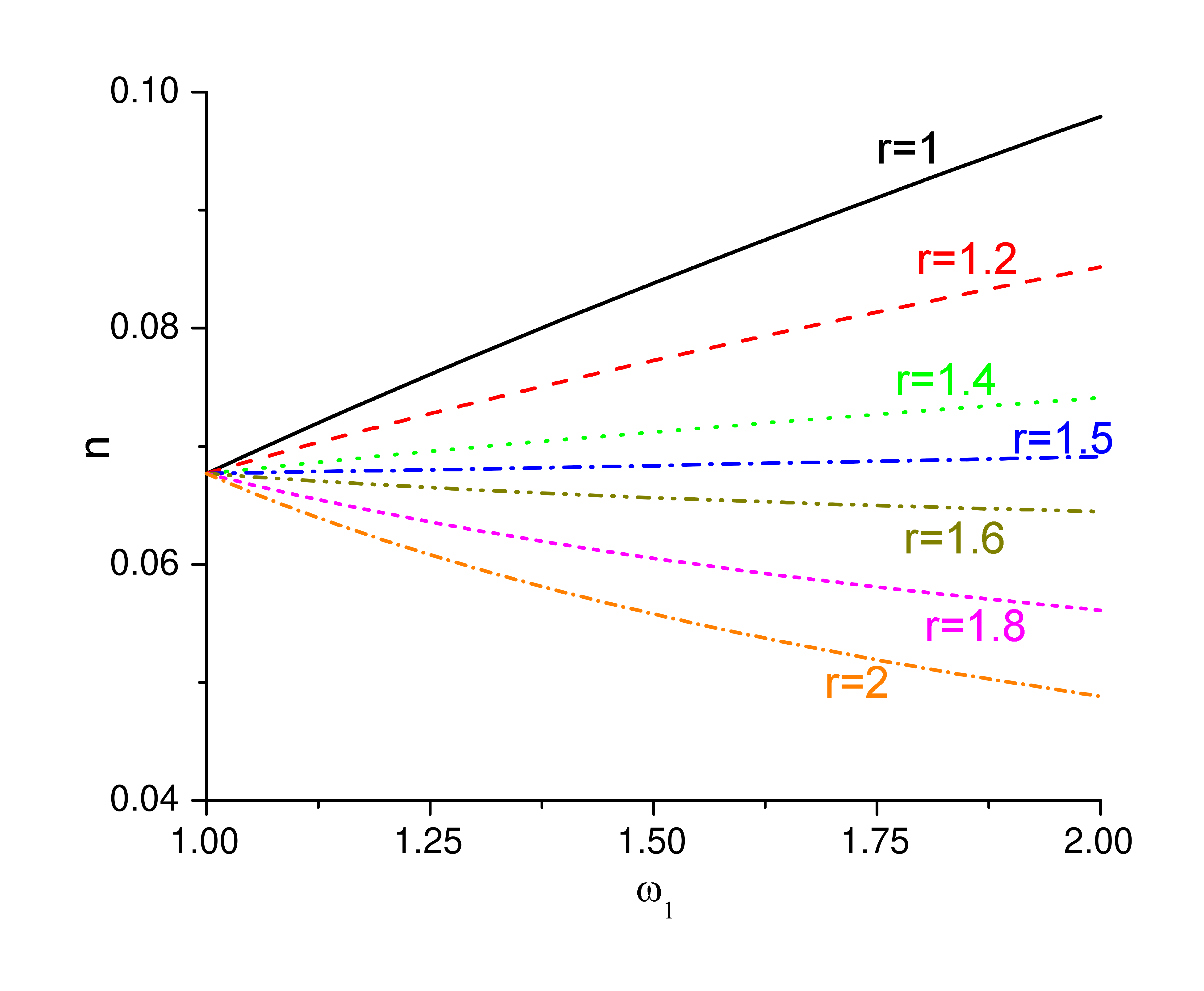}
\end{center}
\caption{Top panels: Plot of $n$ vs $\omega_1$ with $\omega_2=
\omega_1^r$ showing the crossover between regimes with increasing
and decreasing $n$ as a function of $\omega_1$. All parameters are
same as in Fig.\ \ref{fig1}.} \label{fig2}
\end{figure}

From Eq.\ \ref{scaling2}, we also expect that there are two separate
regimes where the behavior of $n$ is qualitatively different when
both $\omega_1$ and $\omega_2$ is increased keeping
$\omega_2=\omega_1^{r}$ with $r\ge 0$. In the first[second] regime
$n$ increases[decreases] with $\omega_1$ and $\omega_2$. The
crossover between these regimes occurs for $\omega_2 =
\omega_1^{r^{\ast}}$ with $r^{\ast}=3/2$ for any $d$. This crossover
is indicated in Fig.\ \ref{fig2}, where $n$ is plotted as function
of $\omega_1$ with $\omega_2=\omega_1^r$. From the plot, we clearly
find that $n$ displays an increasing(decreasing) trend with
$\omega_1$ for $r<(>)r^{\ast}$. Interestingly, at $r=r^{\ast}$, $n$
becomes independent of $\omega_1$ and $\omega_2$.

Next, we generalize Eq.\ \ref{scaling2} for non-linear ramps for
which the exact solution no longer holds as follows. Consider a ramp
for the Hamiltonian $H_{\bf k}(t)$ (Eq.\ \ref{ham1}) with
$\lambda_1(t)= (\omega_1 t)^{\alpha}$ and $\lambda_2(t)= (\omega_2
t)^{\beta}$ where $\alpha$ and $\beta$ are positive real numbers.
Following Refs.\ \cite{kz1,dam1}, we note that the system enters the
impulse region where excitation production occurs around $t'_{0{\bf
k}_0} = b_{{\bf k}_0}^{1/\alpha}/\omega_1$. Following Ref.\
\cite{ks2}, we linearize the Hamiltonian around $t=t'_{0{\bf k}_0}$
and ${\bf k}={\bf k}_0$ to obtain
\begin{eqnarray}
H_{k}^{\rm eff}(t) = \alpha b_{{\bf k}_0}^{(\alpha-1)/\alpha}
\omega_1 (t-t'_{0{\bf k}_0}) \tau_3 + (\omega_2 t'_{0{\bf
k}_0})^{\beta} k \tau_1.  \label{effham1}
\end{eqnarray}
The defect production for dynamics governed by $H_{k}^{\rm eff}(t)$
can be easily found; one can read off the off-diagonal element as
$\Delta = (\omega_2 t'_{0{\bf k}_0})^{\beta} k = (\omega_2 b_{{\bf
k}_0}^{1/\alpha}/\omega_1)^{\beta} k$ and $dE_{\bf k}(t)/dt \simeq
\alpha b_{{\bf k}_0}^{(\alpha-1)/\alpha} \omega_1$. This allows one
to obtain $p_{\bf k} \sim \exp[-\pi \Delta^2/|dE_{\bf k}(t)/dt|]
\sim \exp[-\omega_2^{2\beta} \omega_1^{-(2\beta+1)} k^2]$ which
leads to
\begin{eqnarray}
n \sim \omega_1^{(2 \beta +1)d/2} \omega_2^{-\beta d}, \quad Q \sim
\omega_1^{(2\beta +1)(d+1)/2} \omega_2^{-\beta(d+1)}.
\label{scaling3}
\end{eqnarray}
Note that Eq.\ \ref{scaling3} reproduces Eq.\ \ref{scaling2} for
$\beta=1$ and the standard one parameter drive scaling relations
\cite{kz1,pol2,ks2} for $\beta=0$.

\begin{figure}[tbp]
\begin{center}
\includegraphics[angle=0,width=0.5\textwidth]{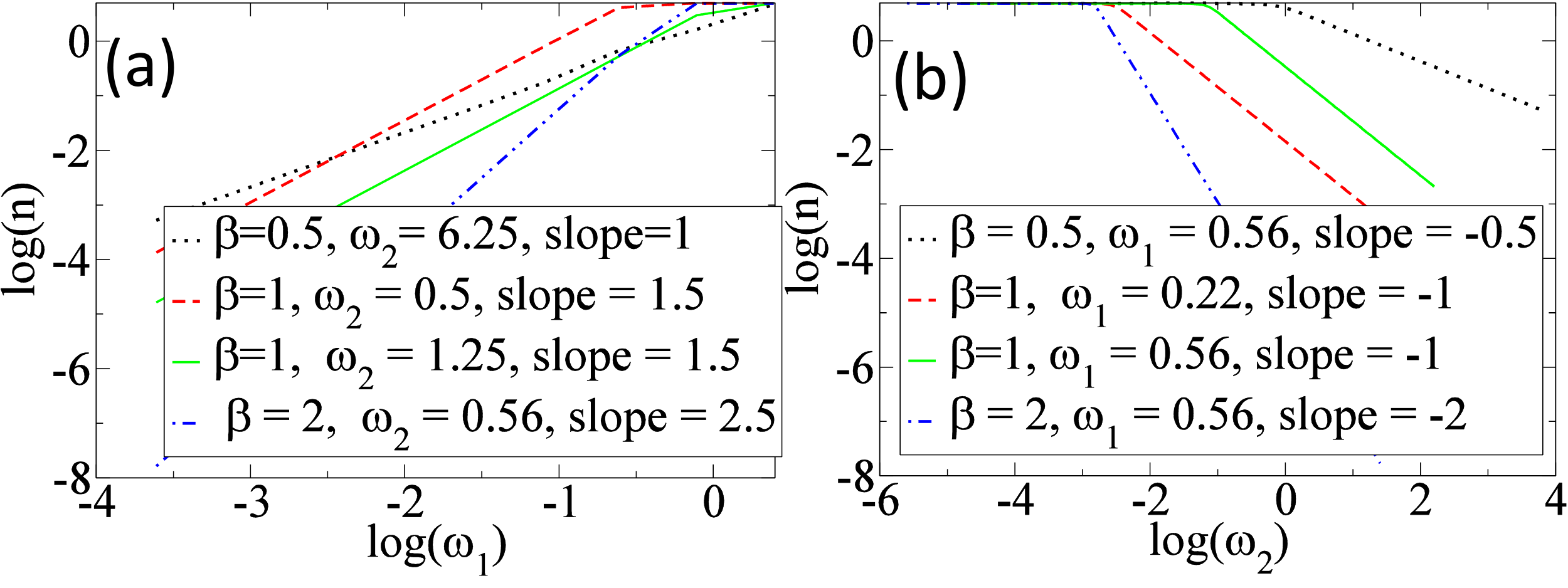}
\end{center}
\caption{ Color online: Plot of $\ln(n)n$ vs $\ln(\omega_1)$ (left
panel) and $\ln(\omega_2)$ (right panel) showing slopes that are
consistent with the theoretical exponents $(1+2\beta)d/(2z)$ and
$-\beta d/z$ respectively (Eq.\ \ref{scal1}). All parameters are
same as in Fig.\ \ref{fig1}. The exponent $\alpha$ does not appear
to influence the slope and is taken to be $2$. The frequencies
$\omega_2(\omega_1)$ are held fixed as indicated in the left(right)
panels.
 } \label{fig3}
\end{figure}

Eq.\ \ref{scaling3} can also be verified by a rigorous analysis. To
this end, we note that the Schrodinger equation corresponding to the
Hamiltonian given in Eq.\ \ref{effham1} can be cast in the form of a
Bloch equation
\begin{eqnarray}
\partial_{t'}{\bm s}=\bm B(t')\times \bm s(t'),\label{bloch1}
\end{eqnarray}
where $\bm B(t')=[(\omega_2 t'/b_{\bf k})^\beta g_{\bf k}/b_{\bf
k},0,(\omega_1 t'/b_{\bf k})^\alpha/b_{\bf k}-1]$ is an effective
magnetic field corresponding to the Hamiltonian $H_{\bf k}(t)$ (Eq.\
\ref{ham1}) and we have scaled $t'= b_{\bf k} t$. The spin variable
$\bm s(t')=\expect{\psi(t')|\bm \tau|\psi(t')}$ in the Bloch
equation characterizes the solution $\ket{\psi(t')}$ of the
Schrodinger equation $i\partial_{t'}\ket{\psi(t')}=H_{\bf
k}(t')\ket{\psi(t')}$ and $\bm\tau=\tau_1 \hat{\bm x}+\tau_2\hat{\bm
y}+\tau_3\hat{\bm z}$ is the vector of Pauli matrices. For future
convenience, we introduce the rescaled frequencies
$\tilde{\omega}_2=\omega_2 g_{\bf k}^{1/\beta} b_{\bf
k}^{-(1+\beta)/\beta}$ and $\tilde{\omega}_1=\omega_1 b_{\bf
k}^{-(1+\alpha)/\alpha}$; in terms of these rescaled frequencies,
one can write
\begin{equation}
\bm B(t')=[(\tilde{\omega}_2 t')^\beta,0,(\tilde{\omega}_1
t')^\alpha-1].
\end{equation}
Next, we rewrite $\tilde{\omega}_2$ in terms of a new variable
$\Gamma$ so as to make the adiabatic limit $\tilde{\omega}_1
\rightarrow 0$ more transparent. To this end, the LZ impulse region
occurs where $t'\sim 1/\tilde{\omega}_1$. In this region, the
parameter which controls the transition probability is
\begin{eqnarray}
\Gamma&=& \left(
|B(\tau)|^2/(2|\dot{B}(\tau)|)\right)|_{\tau=\tilde{\omega}_1^{-1}}
=\tilde{\omega}_2^{2\beta}/(2\alpha\tilde{\omega}_1^{1+2\beta}).
\nonumber\
\end{eqnarray}
We will therefore parametrize $\tilde{\omega}_2$ as
$\tilde{\omega}_2=(2\alpha\tilde{\omega_1}^{1+2\beta}\Gamma)^{1/2\beta}$.
The transition probability $\Phi$, which is a function of
$\tilde{\omega}_1,\tilde{\omega}_2$ is then more conveniently
expressed as a function  $\Phi(\tilde{\omega}_1,\Gamma)$ since it
clearly reproduces the adiabatic limit where it is determined by
$\Gamma$ alone. In terms of this scaling function $\Phi$, $n$ can be
written, as
\begin{align}
&n = \int_0^{\Lambda^{(0)}} \frac{d^d k}{2\pi} \Phi\Big(\omega_1
b_{\bf k}^{-(1+\alpha)/\alpha}, \frac{\omega_2^{2\beta} g_{\bf
k}^{2}b_{\bf k}^{(2\beta-\alpha+1)/\alpha}}
{2\alpha\omega_1^{1+2\beta}} \Big), \nonumber\
\end{align}
where we have expressed $\Gamma$ in terms of $\tilde \omega_2$ and
$\Lambda^{(0)}$ represents the finite range of momentum integration
that can be set of $\Lambda^{(0)}\rightarrow\infty$ in the scaling
limit where $\omega_{1,2}/b_{\bf k_0}^{(1+\alpha)/\alpha} \ll 1 $
and $\omega_2/\omega_1 \ge [ \omega_1 \alpha b_{\bf k_0}
^{(\alpha-1)/\alpha}]^{1/(2\beta)} /b_{\bf k_0}^{1/\alpha}$
\cite{comment1}. In the limit $\omega_1\rightarrow 0$, the integral
is dominated by ${\bf k}\sim {\bf k_0}$ such where $g_{\bf k} \sim
k^z$ and $b_{\bf k}\rightarrow b_{\bf k_0}$. In this case, it is
convenient to rescale the integration variable $k=
[2\omega_1^{1+2\beta}\alpha k'/(\omega_2^{2\beta}b_{\bf
k_0}^{(2\beta-\alpha+1)/\alpha})]^{1/2z}$ so that one can write
\cite{comment2}
\begin{align}
&n \sim
\left(\frac{2\omega_1^{(1+2\beta)}\alpha}{\omega_2^{2\beta}b_{\bf
k_0}^{(2\beta-\alpha+1)/\alpha}}\right)^{d/2z} \int_0^{\infty}
\frac{dk'}{2 z} k^{'\,d/2z-1} \Phi(0,k'), \label{scal1}
\end{align}
which confirms the scaling relation (Eq.\ \ref{scaling3}) for $z=1$.
A plot of $\ln(n)$ obtained by direct numerical solution of the
Schrodinger equation corresponding to $H_{\bf k}(t)$ (Eq.\
\ref{ham1}) with $\lambda_1(t)=(\omega_1 t)^{\alpha}$ and
$\lambda_2(t)=(\omega_2 t)^{\beta}$ as a function of $\ln(\omega_1)$
(left panel) and $\ln(\omega_2)$(right panel) for $d=z=1$,
$\alpha=2$, $b_{\bf k}=5-\cos(k)$, and $g_{\bf k}=\sin(k)$ and
several values of $\beta$, shown in Fig.\ \ref{fig3}, also confirms
these scaling relations.
%Note that in the adiabatic limit, $\Phi(\tilde{\omega}_1\rightarrow
%0,\Gamma)\rightarrow e^{-\pi\Gamma}$, and one obtains
%\begin{align}
%&n_{ad}(\omega_1,\omega_2)=\left(\frac{2\omega_1^{(1+2\beta)}\alpha}{\omega_2^{2\beta}b_0^{(2\beta-\alpha-1)/\alpha}}\right)^{d/2z}\pi^{-(d-1+1/2z)}\nonumber\\
%& \times \Gamma(d-1+1/2z)\Gamma_{inc}(d-1+1/2z,\Gamma_{max}\pi),
%\end{align}
%where $\Gamma_{inc}(a,b)$ is the incomplete Gamma function.

Next, we provide a general system-independent scaling argument which
leads to Eq.\ \ref{scaling1}. We consider a generic Hamiltonian with
two tunable parameters which are varied with rates $\omega_1$ and
$\omega_2$. The first parameter $\lambda(t)$ controls the distance
of the system from a quantum critical point at $\lambda=\lambda_c
\ne 0$; for a generic Hamiltonian, this necessitates that the
instantaneous energy gap near the critical point varies as
$\Delta({\bf k}={\bf k}_0;\lambda) \simeq |\lambda(t)|^{z\nu} =
|\omega_1 t -\lambda_c|^{z\nu \alpha}$, where $\alpha$ is a positive
exponent and $\alpha=1$ denotes linear drive protocol. The second
parameter, $c(t)$, controls the dispersion of the quasiparticles at
the critical point so that $\Delta(k,\lambda_c) \simeq c(t) k^z =
|\omega_2 t|^{\beta} k^z$. Since the defect production occurs in the
impulse region, which for small $\omega_1$ is also the critical
region, we first estimate the time spent by the system in this
region. The Landau criterion for the system to be in the impulse
region is given by \cite{pol1} $d \Delta/dt \simeq \Delta^2$.
Substituting the expression for $\Delta({\bf k}_0,\lambda(t))$ in
this relation, one obtain an expression for $T$, the time spent by
the system in the impulse region, as $|T-T_0| \simeq
\omega_1^{-\alpha z\nu /(\alpha z \nu +1)}$, where $T_0=
\lambda_c/\omega_1$ is the time at which the system reaches the
critical point. Substituting the expression for $T$ in the
expression for $\Delta({\bf k}_0,\lambda)$, one finds that in the
impulse region, the instantaneous energy gap behaves as
\begin{eqnarray}
\Delta({\bf k}_0;\lambda) \simeq \omega_1^{\alpha z\nu /(\alpha z
\nu +1)} \label{enbehav}
\end{eqnarray}
which is in agreements with its counterpart for single parameter
drive \cite{pol1,pol2,ks2}. Next, we note that the defects or
excitations are typically produced in a phase space $\Omega \sim
k^d$ around the critical mode. For these modes, in the critical
region, and during the time $T$ that the system spends in this
region, one has
\begin{eqnarray}
k &\simeq& |\omega_2 T_0|^{-\beta/z} \Delta({\bf
k}_0,\lambda(T))^{1/z}
\nonumber\\
&=& |\omega_2 \lambda_c/\omega_1|^{-\beta/z} \Delta({\bf
k}_0,\lambda(T))^{1/z}. \label{komegarel}
\end{eqnarray}
Using Eqs.\ \ref{komegarel} and \ref{enbehav}, one finally gets
\begin{eqnarray}
n \sim \Omega \simeq \omega_2^{-\beta d/z}
\omega_1^{\left(\frac{\alpha \nu}{\alpha z \nu +1} +
\frac{\beta}{z}\right) d}.  \label{nfinal}
\end{eqnarray}
which reproduces the first relation in Eq.\ \ref{scaling1}. We note
that for the above arguments to hold we need near adiabatic dynamics
which requites $\Delta({\bf k}_0;T)\ll \Delta({\bf k}_0;0)$ in the
impulse region leading to $\omega_1 \ll \lambda_c^{\alpha z \nu+1}$.
Further, one also needs excitation production to occur at the
neighborhood of $k \to 0$ which occurs when $(d\Delta
(k,\lambda_c)/dk^z|_{t=T_0})^2 \ge d \Delta({\bf
k_0};\lambda_c)/dt|_{t=T}$ and leads to the condition
$\omega_2/\omega_1 \ge \omega_1^{\alpha \nu/[\beta(\alpha z \nu
+1)]}/\lambda_c$. Also, the present analysis provides a general
physical understanding of the defect suppression with increasing
$\omega_2$; it occurs due to the reduction of available momentum
modes for quasiparticle excitations at any given energy
$\Delta(k;\lambda)$ with increasing $\omega_2$. Thus the role of the
drive protocol changing $c(t)$ is to reduce the available phase
space for defect production which naturally leads to suppression of
$n$ and $Q$ with increasing $\omega_2$. The expression for the
residual energy $Q$ can be similarly obtained by noting that the
energy of the excitations produced for any $k$ is given by $E[k]
\simeq {k}^z$. This leads to $Q \sim k^z \Omega \sim k^{d+z} \sim
\omega_2^{-\beta (d+z)/z} \omega_1^{\left(\frac{\alpha \nu}{\alpha z
\nu +1} + \frac{\beta}{z}\right) (d+z)}$. From Eq.\ \ref{scaling1},
we also find that the crossover between the regimes where $n$
increases/decreases with $\omega_1$ occurs for $\omega_2 =
\omega_1^{r^\ast}$ with ${r^\ast}= 1+ \alpha z \nu/(\beta(\alpha z
\nu +1))$ which reduces to the condition $r^{\ast}=3/2$ derived
earlier for $\alpha=\beta=z=\nu=1$.

Finally, we discuss concrete models where our theory shall hold.
First we note that for $d=1$, Eq.\ \ref{ham1} represents the XY
model in a transverse field with the identification: $b_{\bf k}= h_0
-\cos(k)$, $g_{\bf k}=\sin(k)$ $ J_x[J_y]=1/2[1+(-)\omega_2 t]$ and
$h(t)= \omega_1 t-h_0$. Our analysis leading to Eqs.\ \ref{exexp}
and \ref{scaling2} is therefore directly applicable to this model
which has been extensively used in the past for test bed for
Kibble-Zureck (KZ) scaling \cite{pol1,dziar1,dutta1,dutta2}. Second,
ultracold superfluid fermions with tunable zeeman field and
spin-orbit coupling is another example where our scaling analysis is
expected to be relevant. The effective Hamiltonians for such
fermions is given by \cite{supp1}
\begin{eqnarray}
H_{\rm eff} &=& \sum_{\bf k} \psi^{\dagger}_{\bf k} \left[ \tau_z
(v(t)-v_{zc}) + \tau_x \alpha(t) g(k) \right] \psi_{\bf k}
\end{eqnarray}
where $v(t)=\omega_1 t$ is the tunable zeeman field, $v_{zc}=
\sqrt{\Delta_0^2 +\mu^2}$, $\Delta_0$ is the superfluid order
parameter, $\mu$ is chemical potential, $\alpha(t)=\omega_2 t$ is
the amplitude of spin-orbit (Rashba) coupling, and $g(k)=\sin k$.
The analysis leading to Eqs.\ \ref{exexp} and \ref{scaling2}
directly holds for this system. In addition, it has the advantage of
being easily implementable using ultracold fermion systems. Finally,
we note that almost all quantum systems near a phase transitions can
be described a Landau-Ginzburg action which has the generic form
\begin{eqnarray}
S &=& \int d^dr dt  \psi^{\ast} [ - \partial_t^2 + c_1 \sum_{i=1,d}
\partial_{x_i}^{2z} + (r-r_c) - u |\psi|^2 ] \psi \nonumber
\end{eqnarray}
Here $r$ controls the distance to criticality while $c_1$ controls
the quasiparticle dispersion at criticality. Our analysis holds for
such theories if $r$ and $c_1$ is tuned as functions of time with
rates $\omega_1$ and $\omega_2$. These parameters are derivable, in
principle, from the microscopic parameters of the system action;
thus our method provides a generic algorithm for defect suppression
by tuning microscopic parameters of a quantum system. We recognize
that the precise experimental implementation, found to be simple for
specific systems discussed above, could be difficult for generic
actions (specially for strongly interacting systems where relation
between microscopic tunable parameters and $r,\,c_1$ may be
complicated); however, the present analysis at least serves as the
first guideline in this respect.

In conclusion, we have obtained novel scaling laws for $n$ and $Q$
for a quantum system driven through a critical point with two rates
$\omega_1$ and $\omega_2$ for arbitrary power law protocols. Our
results constitute a generalization of KZ scaling to two parameter
drive protocols. These results indicate suppression of both $n$ and
$Q$ with increasing $\omega_2$ and therefore provides a route to
shortcut to adibaticity for driven quantum critical systems.

KS thanks D. Sen for discussions and S. Das Sarma for hospitality
during early stages of this work. We also thank I. Spielman and 
F. Setiawan for discussions on a related work that motivated the 
present results. JS thanks the JQI/PFC, CMTC and the University of 
Maryland for start-up support.

\section{Supplementary Material: Spin orbit coupled cold atomic gases}

The Bogoliubov de-Gennes Hamiltonian for a spin-orbit coupled cold
atomic fermi gas with attractive interactions is written as
\begin{align}
&H_{BdG,k}(t)=((\epsilon_k-\mu(t)+\alpha(t)
k\sigma_z)\tau_z+V_Z(t)\sigma_x+\Delta\tau_x,
\end{align}
where $\alpha(t)$ is the Rashba spin-orbit coupling, $V_Z(t)$ is the
Zeeman splitting and $\mu(t)$ is the chemical potential. The
dispersion $\epsilon_k=k^2$ where the effective mass has been set to
$m=0.5$ by an appropriate choice of units. In the experimental
set-up involving artificial gauge fields \cite{Ian1}, the Rashba
spin-orbit parameter $\alpha(t)$ is set by the angle of the incident
Raman beams and the Zeeman potential is controlled by the intensity
of the Raman beams. The matrix $\sigma_{x,y,z}$ represent the spin
degree of freedom, while the matrices $\tau_{x,y,z}$ represent the
particle-hole degree of freedom.

In the adiabatic limit of slow frequency, transitions only occur
near the critical gap closing point of the Hamiltonian
$H_{BdG,k}(t)$ for $t$ such that
\begin{align}
&\Delta^2+\mu(t_c)^2=V_{Z}(t_c)^2
\end{align}
and near $k\approx 0$. Near the transition point, one can ignore the
$\epsilon_k\sim k^2$ term and approximate the Hamiltonian as
\begin{align}
&H_{BdG,k}(t)\approx \alpha(t)
k\sigma_z\tau_z+V_Z(t)\sigma_x+\Delta\tau_x-\mu(t)\tau_z.
\end{align}
Focusing on $k=0$, we note that only one pair of eigenstates
$|{\pm}\rangle$ of $H_{BdG,k=0}(t\sim t_c)$ are at energies of order
$t\sim t_c$, while other pair are at energy
$\sqrt{\Delta^2+\mu(t_c)^2}$. Projecting at small $k$ into these
eigenstates
\begin{align}
&H_{BdG,k}(t)\approx \alpha(t) k[|-\rangle \langle +|\langle
+|\sigma_z\tau_z|-\rangle +h.c]
\nonumber\\
& +[V_Z(t)-\sqrt{\Delta^2+\mu(t)^2}][|+\rangle\langle +| - |-\rangle
\langle -|].
\end{align}
Defining a pseudo-spin $\rho_x=|+\rangle \langle-| +h.c$ and
$\rho_z=[|+\rangle \langle +| - |-\rangle \langle -|]$, we can write
the effective Hamiltonian as
\begin{align}
&H_{BdG,k}(t)\approx
k\alpha(t)\frac{\Delta}{\sqrt{\Delta^2+\mu(t)^2}}\rho_x
\nonumber\\
& +[V_Z(t)-\sqrt{\Delta^2+\mu(t)^2}]\rho_z.
\end{align}
The chemical potential $\mu(t)$ in a closed system should be set by
the density. In the limit of small $\Delta,V_Z(t),\mu(t)\ll
\alpha(t)$, one can ignore the time-dependence of $\mu(t)$ and
approximate $\mu(t)\approx 0$. This leads to
 \begin{align}
&H_{BdG,k}(t)\approx k\alpha(t)\rho_x+[V_Z(t)-\Delta]\rho_z.
\end{align}
Choosing appropriate time-dependence for $\alpha(t)$ and $V_Z(t)$
leads to the model (Eq.~[17] in the main text) discussed in the main text.

\end{document}